# Wavelet-based filter methods to detect small transiting planets in stellar light curves

Sascha Grziwa[1] & Martin Pätzold[1]

[1]Rheinisches Institut für Umweltforschung an der Universität zu Köln

Abt. Planetenforschung

Aachenerstrasse 209, 50931 Cologne,

Germany, (email: grziwas@uni-koeln.de)

**Abstract**

Strong variations of any kind and causes within a stellar light curve may prohibit the detection of transits, particularly of faint or shallow transits caused by small planets passing in front of the stellar disk. The success of future space telescopes with the goal for finding small planets will be based on proper filtering, analysis and detection of transits in perturbed stellar light curves. The wavelet-based filter methods VARLET and PHALET, developed by RIU-PF, in combination with the transit detection software package EXOTRANS allow the extraction of (i) strong stellar variations, (ii) instrument caused spikes and singularities within a stellar light curve, (iii) already detected planetary or stellar binary transits in order to be able to search for further planets or planets about binary stars. Once the light curve is filtered, EXOTRANS is able to search efficiently, effectively and precisely for transits, in particular for faint transits.

# 1. Introduction

The concept for the detection of transits in stellar light curves is simple. During the transit of a planet in front of a stellar disc, as seen from an observer on Earth or in Earth orbit, the observed stellar flux drops in a characteristic way. The typical shape and periodic appearance of a transit event is used for the detection. It also characterizes the eclipsing object. Real stellar light curves, however, show many flux variations in addition to the transit event. This may cover the planetary transit signal and, in particular, faint transits from small orbiting planets may become undetectable.

The change in stellar flux during a transit event is very small (1% for hot Jupiters; < 0.01% for earthlike planets). The recording of stellar light curves from ground is difficult and challenging because of day/night changes, bad weather, a low signal-to-noise ratio (SNR) caused by atmospheric seeing or pollution by lunar light.

Continuous and high resolution light curves not influenced by the atmosphere or day/night changes are available since the launch of the space telescopes CoRoT (Baglin et al., 2003) and Kepler (Borucki et al., 2010). These two space telescope missions have monitored star fields of 100,000 stars in total and have the capability to detect planets as small as the size of Mercury (Barclay et al., 2013). More than 275 new planets and more than 5000 planetary candidates still pending for confirmation by follow-up observation were discovered during both missions. Future missions like TESS (Ricker et al., 2011) and PLATO (Rauer et al., 2014) will monitor more than 1,000,000 stars and will cover nearly the complete sky for the search for earthlike planets.

Even after eliminating the interference of the atmosphere, the intensity of starlight over time is anything but constant. Variations in the intensity of the light curve are caused by stellar rotation (star spots), pulsations and stellar activity. This composition of different periodic and non-periodic variations are the major reasons for the non-detection or misinterpretation of transits. Discontinuities within the light curves generated by technical errors and saturation of the CCD by impacts of charged particles can also lead to false detections. Even technically more advanced instruments planned for future missions like PLATO will face such problems. Therefore, not only the instruments but also the processing and detection software needs to be improved.

An automated detection procedure is required in order to search the large number of light curves. This can be archived in two steps:

1. The filtering of the light curve by separating varying and perturbing contributions.
2. The subsequent search and detection, which scans the light curve for periodic characteristic transit signals.

The detected transit features are then analyzed and as many system and candidate parameters as possible are determined. There are different individual methods in use (Cabrera et al., 2012, Grziwa et al., 2012, Jenkins et al., 2010, Moutou et al., 2005), with emphasis either on filtering or detection.

One of the most commonly used methods is the combination of filters applied to the light curve with the subsequent search for transits with the Box-Least-Squared algorithm (BLS) (Kovács et al., 2002) or a related algorithm (Moutou et al., 2005, Cabrera et al. 2012). The Rhenish Institute for Environmental Research, Department of Planetary Research (RIU-PF), as part of the CoRoT detection team uses a combination of trend and harmonic filters and a modified BLS algorithm in their detection pipeline EXOTRANS (Grziwa et al., 2012). This detection pipeline was developed for and successfully used during the space mission CoRoT and improved throughout the mission leading to the detection of about 1000 planetary candidates (e.g. LRa01, Carone et al., 2012 for the first long run pointing at the anti-center of the galaxy) in all target star fields. The filter combination is well suited to prepare the light curves for the search with the dcBLS algorithm. The Kepler team relies on a relatively weak filtering of the original light curves and developed a proprietary algorithm for detection (Jenkins et al., 2010).

## 2. Origins of flux variations in light curves

The recorded stellar flux as a function of time (the stellar light curve) contains contributions which may prevent the detection of a transit signal.

Discontinuities in the light curves (Figure 3a) may be caused by temporary technical failures within the CCD device. The Impacts of high energetic particles are single events but cause strong signals. This immediate jump in flux may cause false detections. The extraction or correction of these singularities by the usual filter methods (e.g. high, low and band-pass filters) is very difficult if not impossible.

The dominating limiting factor, however, for the detection of transits is the intrinsic stellar variability of the target star. Stellar variability is independent of the quality of the instrument and the environment of the telescope and limits directly the detection of small planets. Contributions to the stellar flux variability are from star spots, regular and irregular pulsations and non-periodic events like flares. An example of a highly variable light curve caused by stellar activity is shown in Figure 4a.

The variability of the flux has a very complex frequency spectrum and results in very irregular shapes of the light curves. The development of a complex stellar model precise enough to predict the variability over the observation time and as complex as shown in Figure 4a to correct this variability is an impossible task. A model independent filter is needed to correct for the variability in order to be able to resolve the transit feature.

## 3. Transit detection and the need for filters

The BLS algorithm (Kovács et al. 2002; Grziwa et al. 2012) detects 'boxlike' events in stellar light curves. The light curve is phasefolded at a wide range of periods and a square box is fit to the complete phase range. If the light curve is phase-folded at the correct period of the transit event, the transit is amplified several times and the least-squares fit delivers the lowest $\chi^2$. The period with the best fit or lowest $\chi^2$ is chosen as the transit or orbit period.

The fluctuations of stellar variability may produce larger amplitudes then the planetary transit. Without proper filtering before the BLS search, there is a high chance that multi-periodic variability may lead to incorrect detections and/or lead to increased noise in the folded phase-diagram and may prevent the detection of shallow or faint transits. Individual discontinuities are significantly attenuated by the phasefolding. Strong disturbances may be clearly present in the phase-folded light curve and unfortunate coincidences of various disturbances may lead to false detections at individual periods.

The EXOTRANS pipeline was adapted to process light curves of the Kepler space telescope when the first Kepler data for public use became available. RIU-PF was able to detect more than 1000 candidates from the 150,000 light curves from the data of the first Kepler quarter (Q1). A comparison with the official list of candidates of the Kepler team from 2011 showed that RIU-PF detected 76% of the official Kepler candidates of interest (KOI) and additionally 35 very promising unknown candidates (Grziwa et al., 2011).

Although this was a remarkable success for the RIU-PF processing pipeline, it became clear that strong stellar variability was the main reason for missed candidates. For this reason, RIU-PF developed new filter methods, which are capable to extract stellar variability and discontinuities from the light curves. These new filter techniques are based on wavelets. The used Stationary Wavelet Transform Denoising (SWTD) is advantageous over the commonly used Short Time Fourier Transform (STFT) or similar filtering techniques, which use sine and cosine functions.

## 4. Introduction to Wavelets

The theory of wavelets is described for example in Coifman et al. (1995) and cannot be fully covered in this paper. Wavelets are used since the late 1980s starting with the discovery of "orthogonal wavelets with compact support" by Ingrid Daubechies (Daubechies, 1992) and the development of the Fast Wavelet Transform (FWT) by Stéphane Mallat and Yves Meyer (Mallat, 1989). The Fast Wavelet Transform represents a generalization or evolution of the Short Time Fourier Transform.

In Short Time Fourier Transform sine and cosine functions  are used to represent or analyze the target function (light curves in our case). In the Fast Wavelet Transform wavelets are used for transformation instead of sine and cosine functions. The different nature of wavelets opens some new possibilities. Many different wavelets with different characteristics exist. Figure 1 shows as an example the Haar wavelet (also called Daubechies wavelet of first order) This is the simplest form of a wavelet and shows the characteristics a wavelet function must fulfill if it is used for Fourier transform techniques. The mean of the Haar function is zero and the euclidic norm is one. Additionally higher order moments of the wavelet function should vanish.

This can be written in following form:

- The mean of the function is zero:
$$\int_{-\infty}^{\infty} \psi(t)dt = 0$$

- The euclidic norm is one:
$$\int_{-\infty}^{\infty} |\psi(t)|^2 dt = 1$$

- Higher order vanishing moments m<M:
$$\int_{-\infty}^{\infty} t^m \psi(t)dt = 0$$

$\psi(t)$ is the wavelet function. Therefore, wavelets can build an orthonormal basis in $L^2(R)$. So you can use Fourier transform techniques and define the continuous wavelet transform (CWT):

$$CWT_x^{\psi}(\tau,s) = \Psi_x^{\psi}(\tau,s) = \frac{1}{\sqrt{|s|}} \int x(t) \cdot \psi^* \left(\frac{t-\tau}{s}\right) dt$$

$x(t)$ is the test signal and $\psi(\tau,s)$ is the wavelet function with the translation parameter $\tau$ and the scaling parameter s. By using the definition of the inner product in $L^2$ one can rewrite the equation to a simpler form:

$$CWT_x^{\psi}(\tau,s) = \Psi_x^{\psi}(\tau,s) = \int x(t) \cdot \psi^* \, dt$$

where

$$\psi_{\tau,s} = \frac{1}{\sqrt{s}} \psi \left(\frac{t-\tau}{s}\right)$$

The definition of the CWT analog to the Fourier transformation shows that the calculated coefficients can measure the similarity between the wavelets and the signal at a given time of the signal by calculating the coefficients with different translation and scale. Because many wavelets are not defined by functions but by discrete numerical tables the technique is ideally used for applications in software algorithm. Any function can be represented by a numerical table of the wavelet and the corresponding coefficients. If one wants to use wavelet transformation for spectral analysis or filtering some additional characteristics of a wavelet get important. To analyze and represent signals we need wavelets that are local in space and frequency. Typically, this is achieved by building wavelets which have compact support (localization in space), which are smooth (decay towards high frequencies), and which have fast vanishing moments (decay towards low frequencies).

(Sweldens 1995). This opens the chance to invent techniques to analyze or decompose signals which use high frequency resolution in one part and high time resolution in other parts if needed. As usual this is limited by the Nyquist-Shannon sampling theorem (Shannon, 1949) which does not allow infinite frequency and time resolution at the same time. Using classical Fourier transform techniques one must select between high frequency resolution or high time resolution.

Therefore, signal processing is a strong domain for the use of wavelets. A time series, which is composed of various signals with different frequencies changing over time may be represented well by wavelets. Steep flanks (a typical example for fast changing signals) may be represented at a high time resolution without losing the frequency resolution in the remaining part of the light curve.

This makes wavelets the ideal choice to represent complex signals like highly variable stellar light curves. VARLET uses the Stationary Wavelet Transform Denoising (SWTD) to decompose a complex signal using wavelets.

## 4.2   Stationary Wavelet Transform Denoising (SWTD)

Stationary-Wavelet-Transform-Denoising (SWTD) is a technique to reduce noise in complex time series. The discret wavelet transform (DWT) is used to create a set of coefficients which analyzes the frequencies of the different parts of the signal. The coefficients are used to separate the signal with different low- and high-pass filters.
Figure 2 illustrates the single steps to denoise a quadchirp-signal overlaid with white noise (upper left panel). In this example the signal is decomposed in six steps (a5, d1-5). Panel a5 shows the calculated approximation coefficients after the wavelet based first low pass filter was applied.

The approximation coefficients are calculated by building the inner product between the signal and the translated (position on x-axis) and scaled (frequency and amplitude) wavelets. Simultaneously the approximation coefficients of the corresponding high pass filter are calculated. The high pass filtered part of the signal is used for further processing. At the next step the filter process is repeated with higher resolution because lower frequencies were filtered in the last step and have not to be analyzed again (d5 left panel).

The noise limit is calculated (dotted line in the d5 left panel). Only wavelet coefficients exceeding the noise limit in amplitude and frequency are separated (d5 right panel). This filter process is repeated several times with the remaining signal set (d5-d1). Each time a new noise level and a higher wavelet resolution is calculated to separate parts of the main signal from the noise. This results in a number of band pass filters with different selective bandwidths depending on the noise threshold. At the end of the estimation process the different denoised signal sets of different bandwidths (a5, d5-d1 right panel) are combined to give the best wavelet representation of the original complex signal but at a lower noise level (upper right panel). The residual is calculated by subtracting this de-noised signal from the unfiltered signal.

In this example parts of the original signal remain in the filtered residual (sample 100-200). The quality of the filter depends strongly on the selected noise limits and the number of iterations.

## 5. VARLET – noise extraction

### 5.1 Searching for the transit with wavelets, but how?

A planetary transit appears in the light curve as a boxlike event (steep flanks) with shallow depth (0.01%-1%) and short duration (couple of hours). The flux returns to the original level after the transit. The amplitude of stellar variations, however, generally changes more slowly but stronger compared to the fast change of flux of a planetary transit. Many stellar variations exceed strongly the depth of a transit. The flanks of discontinuities caused by instrument errors may be as steep as the flanks of the transits but the amplitude change of discontinuities also exceed the depth of a transit. Neither the frequency nor the amplitude alone can distinguish stellar variation or discontinuities from transits, but the combination can.

SWTD is able to filter noise of a selected or calculated noise level by reconstructing the complex structure of signals with a high number of scaled wavelets. How can this be used to separate stellar variability and discontinuities from shallow transits? The filter routine VARLET, developed by the authors, reconstructs the stellar variability and discontinuities of the original light curve, but does not repeat the fast and relative small changes of a transit.

SWTD is the sequential application of a series of filters. A very high noise threshold is chosen for the first filter. The selected filter frequency shall be lower than the duration of a typical transit. The scaled wavelets used for the first adaption of the light curve do not reconstruct the transit. The noise is separated from the original light curve together with the transit by subtracting the reconstructed light curve from the original light curve. The 'slow' changing stellar variability is roughly reconstructed by the scaled wavelet. These rough approximations are then improved step by step by smaller wavelets which give a better representation of the reconstructed shape of the stellar variability and complex discontinuities. The transits are already separated with the first step.

The reconstructed light curve is a very precise representation of the stellar variability and discontinuities after 20 iterations but does not include transits with durations and amplitudes lower than the scale selected for the first wavelet. The reconstructed light curve is now subtracted from the observed light curve. The residual light curve contains the transit signal and the white noise. EXOTRANS then searches for transits by applying the BLS algorithm and is now able to easily detect even shallow transits once the large-scale stellar variation and discontinuities were removed.

## 5.2   Separation of stellar variability by VARLET

Some examples of light curves with high variability shall be discussed in the following in order to demonstrate the capability of VARLET.

Figure 4a shows exemplarily a highly variable Kepler light curve (KIC 4827723). The wavelet based representation of the stellar variability by the VARLET filter is shown in Figure 4b. The reconstruction of the filter gives a good representation of the complex combination of the strong variability of the light curve. The residual light curve after subtracting the reconstructed light curve (Figure 4b) from the observed light curve (Figure 4a) is shown in Figure 4c. The scale of Figure 4c is zoomed by a factor of 100.

The residual light curve in Figure 4c was searched by EXOTRANS using the dcBLS algorithm (Grziwa et al. 2012). A transit with a period of 7.239 days was detected (Figure 5a).The proper transit of the planetary candidate is easily identified in the phase diagram Figure 5b shows the phase diagram of the same light curve without being filtered by VARLET but also phase-folded at 7.239 days. Stellar variability disturbs strongly the phase diagram.  The detection of the transit is difficult if not impossible.

## 5.3   Separation of discontinuities by VARLET

Figure 3a shows the observed CoRoT light curve CoRoT-id 102586290 (or LRa06 E2_4780). Many discontinuities are identified in the light curve. Figure 3b is the reconstructed light curve. Even fast abrupt changes of the intensity are precisely reconstructed by VARLET. Again, the residual light curve (Figure 3c) is now searched for transits by EXOTRANS. Discontinuities are not contained in the residual light curve and the risk for false detection is minimized compared to the observed original light curve. Figure 3d shows the successful detection of a transit at a period of 6.2182 days.

## 5.4   Transit depth after filtering

Tests have shown that the transits depth is altered by the VARLET filtering process (Figure 6). Because the depth of the transit is an important parameter for the characterization of the candidate (Mandel and Agol, 2002), the residual light curve shall be used for search and detection but the original light curve for the characterization of the transit. (Figure 6).

Nevertheless, after the detection of a transit in the residual light curve, and with the knowledge of the detected period and the mid-transit times, most transits may then be easily identified in the original light curve.

The comparison of the ingress and egress times between the residual and the original light curves show that transit mid-times are not altered by the filter process. (Figure 6a and 6c).

## 6. PHALET - Extraction of transits and the multi-planet search

### 6.1 PHALET filter

Observed stellar light curves may include periodic disturbances at well-known periods like disturbances from the orbit of the space-telescope or background diluting binaries. Narrow band-pass filters like the harmonic filter in EXOTRANS are used to subtract these disturbances (Grziwa et al., 2012). These filters are based on FFT techniques, which use sinusoidal functions to reconstruct the frequency range of the disturbance. Disturbances of complex shape, however, in particular steep flanks are not properly removed by the filter and the residual light curve is strongly altered.

In order to search for transits from multiple planets in the same light curve or from planets in stellar binary systems, the wavelet-based PHALET filter was developed by the authors which extracts transits at a well-determined period.

PHALET uses the SWTD in combination with phase-folding and is able to extract complex events like transits or orbit disturbances at a certain fixed period.

A single planet or stellar eclipsing binary, however, appears in a power spectrum not only with the real orbital period but also with harmonic multiples. If the orbital period of a second planet is in resonance with the orbit of the first eclipsing object, the harmonic multiples of the first eclipsing object are covered by the periods of the second planet in the frequency power spectrum. This makes it extremely difficult to detect that second planet.

One solution to this problem is to search the residual light curve with VARLET / EXOTRANS for the planet with the most significant transits first. The transits are fully extracted from the light curve by PHALET after the detection of the first eclipsing object. The residual light curve is then searched again for additional transits by EXOTRANS.

The function of the PHALET filter is described in three steps:

1. The residual light curve is phasefolded with the well-known period after every segment is detrended separately. Only the feature with the exact same period are amplified by the phasefolding. Features at other periods are smeared out and covered in white noise.
2. The phasefolded light curve is filtered by SWTD to reduce the noise and to reconstruct the structure of the periodic feature.

3. The phasefolded, noise filtered light curve is unfolded and subtracted from the raw light curve.

## 6.2   Example: Subtracting a binary from a light curve

Planets in a stellar binary system are of special interest because of unsolved questions concerning origin, evolution and orbit stability. PHALET is a tool which can be used to filter eclipsing stellar binaries from light curves in order to detect planets in binary systems. The transit of an eclipsing binary is very prominent in a light curve (Figure 7a), and will mask the detection of transits from additional planets in that system.

Simulations of binary system transits widely used for characterization (e.g. DEBIL; Devor, 2005) cannot be used to subtract the transit of the stellar binary. These models cannot simulate the complex structure of the eclipsing binary transit (reflection and tidal effects) in the light curve precisely enough to subtract the transit completely. Strong artifacts remain that disturb the detection of additional transits.

PHALET reconstructs the structure of the transit by using only the precisely detected period. Figure 7a shows the Kepler light curve KIC 4474645 phasefolded and binned at the detected period of the eclipsing binary transit (3.89 days).

Figure 7b shows the residual light curve at the detected period after PHALET filtering and binning. PHALET removed completely the transit of the stellar binary without a-priori knowledge or applying any model. No artefacts of the binary transit are visible after reducing the noise by binning. A search for additional planetary transits with EXOTRANS is now feasible.

## 6.3   Example: multi-planet search

The identification of planets on resonant orbits is difficult because of the overlay of the partial harmonics of the orbital period of both planets in the periodogram. A good example is the discovery of Kepler 36b, and Kepler 36c (Carter et al., 2012). While Kepler 36c (period = 16.23 days) was detected by the transit method, Kepler 36b (period = 13.825 days) was detected by Transit Time Variation (TTV).

The Kepler 36 light curve (Figure 8) shall be used here exemplarily to demonstrate the capabilities of VARLET and PHALET. The original light curve (Figure 8a) was filtered by VARLET to remove all stellar variabilities (Figure 8b). Afterwards, the residual light curve (Figure 8c) was searched for transits by EXOTRANS and Kepler-36c is found (Figure 8d). The residual light curve (Figure 8c) is now filtered by PHALET at the period of Kepler 36c which extracts all transits of Kepler-36c (Figure 8e). EXOTRANS searched now again and Kepler-36b is found in transit (Figure 8f).

Figure 9 shows the resulting EXOTRANS periodogram for the various tested periods in the range from 0.5 days to 20 days. The higher the power value in the periodogram, the higher is the probability of a box like transit event at that associated period. The highest peak value in Figure 9a is found for a period of 16.23 days, which is the orbital period of Kepler 36c. Other high values are harmonics at half the period and other partials of the orbital period of Kepler 36c.

A very weak double peak is also visible at the period of Kepler 36b (13.825 days) in Figure 9a. More than 20 other harmonics, however, show higher values than that of Kepler-36b. A harmonic at the 6:7 resonances of the transit of Kepler-36c overlays in fact the detection peak of Kepler 36b. As a conclusion, the direct detection of Kepler 36b by EXOTRANS would be very unlikely.

The residual light curve was then filtered with PHALET at the orbit period of 16.23 days (Kepler-36c) and searched a second time by EXOTRANS. Figure 9b shows the periodogram after filtering and the second EXOTRANS run. The line at the period of Kepler-36c of 16.23 days disappeared within the noise as well as all partial harmonics of that period. Comparing Figure 9a and 9b, however, the line caused by the orbital period of Kepler 36b is not affected by the PHALET filter.

## 7. Testing the performance of VARLET

A performance test was executed by the CoRoT detection groups (Moutou et al., 2005) before the launch of CoRoT. About one thousand artificial light curves were provided to be processed by the detection groups, each using their own software and detection philosophy. Only 20 of the 1000 light curves contained transits of various kinds. The detection groups, however, did not know which ones. The results of this "blind test" were reported by Moutou et al. (2005). The authors performed a similar "blind test" to improve the performance of EXOTRANS. Real CoRoT light curves with simulated transits were used this time (Grziwa et al., 2012). The best way to test the performance of the VARLET filter is the use of real data with artificially implanted transits. Transits of different depths and duration were added to the light curves which were then first filtered by VARLET and then processed by EXOTRANS for the search of these transits.

A sample of 2000 light curves was randomly selected out of the total ensemble of 150,000 Kepler light curves. Simulated transits were embedded in these real light curves in order to test the performance of VARLET. There was a high probability that many of the selected 2000 light curves eventually contain already transits of real planets. Therefore, the set has been filtered by VARLET and searched for transits by EXOTRANS. A total of 137 light curves with positive transit detection were rejected from this set.

Transit shapes, depth and duration were simulated with the Mandel and Agol algorithm (Mandel and Agol, 2002) for four different kinds of planets (Jupiter, Neptune, super-Earth and Earth) and four different orbital periods (1.53 days, 5.31 days, 10.31 days, and 19.29 days). The strange orbital periods were selected to avoid resonances and overlapping partial harmonics. The spectral classes of the selected stars are known, their masses and radii are listed in catalogues. Finally, the relative planetary radii (which influence the transit duration) were calculated. In the end, 29,778 light curves (4 planets x 4 periods x 1863 light curves) were seeded with simulated transits.

In order to test the performance of VARLET, the set of 29,778 light curves was

i) searched for transits by EXOTRANS without being filtered by VARLET.

ii) first filtered by VARLET and then searched for transits by EXOTRANS.

A positive detection was counted if the detected transit period was in the range ± 0.01 days of the true period of the simulated transit.

## 7.1    Methods to evaluate the quality of the filter and false positives

In general, all detection which are not caused by the transit of a planet are false positives. Possible false positives are Disturbances or jumps, stellar variabilities, binary stars and contaminating binaries. But not all types of false positives are excluded by filter or the detection algorithm. Transits of binaries and especially contaminating binaries are not completely reduced by filtering. Because of the similarity between transits of binaries and planetary transit the BLS algorithm cannot distinguish between binaries and planets. Only the subsequent analysis by binary simulation or software for probability estimation like Pastis (Díaz et al. 2014) or Blender (Torres et al. 2011) helps to distinguish planetary transits from binary transits. Finally, only additional methods like radial-velocity measurements or transit time variation (TTV) can reveal the real nature of the transit. Therefore, this kind of false positives are not included in this filter test. Nevertheless, PHALET can remove these transits to search for additional transits in the light curve.

Unfiltered strong disturbances and periodic variabilities in light curves are sometimes detected with higher significance than faint transits of planets by the BLS algorithm. A "perfect" filter would reduce all these disturbances or variabilities much stronger than the planetary transit. In the filtered light curve the remaining planetary transit is detected by the detection algorithm. In this test scenario all Kepler light curves show different disturbances and variabilities with different strength. Planetary transits were injected in every light curve. Therefore every transit in this test which was not detected with the correct period by the BLS algorithm is a false positive. Disturbances and variabilities were not completely reduced to reveal the transit of the planet.

The BLS-algorithm evaluates the period which is most similar to a transit for each light curve. For each light curve a Signal-Detection-Efficiency value (SDE) is calculated depending on the transit-like event and in relation of the significance of events with different periods in the light curve. Filtering disturbances and variabilities without altering the transit raises the SDE-value of the detected planetary transit. This is important, because only detections with a SDE-value higher than the statistically calculated SDE-threshold (Grziwa et al. 2012) are considered for further analysis. Therefore also the combination of filter and detection algorithm has to examined, by analyzing the SDE-values of the detected planets.

## 7.2   Performance

Table 1 lists the number of detections at the correct orbital period (+/- 0.01 days) by EXOTRANS after VARLET filtering of the light curves and the percentage relative to the set of 1863 light curves.

The following trends are obvious:

i) The number of detections decreases with increasing orbital period, respectively orbital radius.

ii) The number of detections increases with increasing planetary radius for each orbital period.

Small planets produce shallow transits in the light curve and the situation worsens with increasing orbital radius. The number of transits in a specific light curve decreases with increasing orbital period. As a consequence the transit SNR in the phase-folded diagram decreases in a comparable noisy environment and the transit is harder to identify. Only 55 of the short-period Earth-sized planets could be identified, barely 5% at the long orbital periods.  A possible explanation for this low detection rate is the low sampling time (30 minutes) of the short cadence Kepler light curves. A shallow transit produced by a small planet on a longer period orbit is covered by just 5 observation points. Shape, depth and duration of a shallow transit can barely be represented statistically by this small number of points. The SNR is insufficient to resolve the transit from the white-noise background.

The conclusion above is also true for the super-Earth type planets where the detection of the correct orbital period dropped below 50% for periods of 19 days although most of the short-period transits could be identified.

The orbit period of the transits produced by the large planets of Neptune and Jupiter sizes were mostly correctly identified regardless of the orbital period. The explanation is that the SNR of the transit in the phase-folding diagram is so strong that the transit cannot be misidentified.

It has been shown that VARLET is capable to reduce the general variability within a light curve efficiently. Even a detection of transits produced by planets below the Earth's radius is probable if the sampling time is shorter (<10 min) than that used by Kepler (30 minutes).

The sampling time of the Kepler light curves was chosen by the Kepler project such that the detection of a planet with one Earth radius is theoretically just feasible. A transit of an Earth-like planet is then represented by four to five points only. This is the lower theoretical limit to fit a box to the data. It has been shown here that this is statistically insufficient and the probability of none or a false detection of the BLS algorithm especially for large periods is very high. The superior performance of the VARLET filter is in particular demonstrated if the previous results are compared with the detection results of the unfiltered light curves (Table 2). The transits of Earth-, super-Earth and Neptunes are practically not detected. The identification of transits produced by Jupiter-size planets is below 50% for short periods and drops to 5% for long periods. This comparison demonstrates clearly the need for proper filtering of the light curve before the transit search. The VARLET wavelet filter is the high efficiency tool to reach this goal.

Figure 10 examines the combination of filter and detection algorithm regarding a chosen SDE-threshold for the test scenario. Exemplary for the period of 1.51 days the percentage of detected planets against the cumulative SDE-values is shown. In that way the percentage of detected planets for a specific SDE-threshold can be read directly. The dashed line shows the SDE-threshold of 11 which was calculated from the control group. The SDE distributions of Jupiter and Neptune size planets shows very high SDE values (related to the SDE-threshold). This shows that the transit is very significant against the disturbances and variabilities in the light curve. This is also true for super-Earth size planets. The cumulative distribution shows the same shape with slightly lower detection percentage for the SDE values respectively. Nevertheless nearly all detected super-Earth size planets have SDE-values higher than the calculated SDE-threshold while 95% of all light curves have SDE-values lower than the SDE-threshold. In contrast the cumulative SDE-distribution of the detected Earth size planets shows a completely different shape. 45% are not detected and additional 15% have SDE-values lower than the SDE-threshold.

## 8. Summary and Conclusion

The wavelet-based filter routine VARLET was developed to extract stellar variability and discontinuities from high-resolution light curves observed by space telescopes like CoRoT and Kepler, and which shall be used within future space missions like TESS and PLATO.

VARLET can be applied to light curves without a-priori knowledge, any assumptions or modeling of the disturbance. VARLET is capable of processing extremely large data sets like the accumulation of the 16 quarters of the Kepler light curves in one run.

Equidistant sampling is not necessary in contrast to other filter techniques. Offsets and gaps in the data series do not interfere with the performance of the filter. VARLET is fast enough to process the complete Kepler data set within a few days on a typical PC. This makes it an ideal tool to meet the increasing requirements for data processing of future space telescopes.

It was shown that VARLET is able to suppress the variability within light curves by a factor of 100 and thus to improve the detection of faint or shallow transits significantly and efficiently, compared to unfiltered light curves.

Tests have shown that even shallow transits from Earth sized planets can be detected if the photometric sampling time is chosen optimally. The VARLET detection rate may be sufficient to make a statistically significant statement on the occurrence of planets from the size of super-Earths-like planets and upwards.

It is strongly recommended to increase the photometric sample rate for future space missions than the theoretically required sampling rate for detection if technically feasible.

PHALET is a second wavelet-based filter routine. This filter is able to remove events with complex shapes at well-known periods. This filter is applied to the multi-planet search. Transits from detected planets or transits from binary stars are removed from the residual light curves. Afterwards these light curves are searched again for further transits by EXOTRANS. Also planets near orbital resonances are easily detected.

In a follow-up paper, we shall present the results of the VARLET/PHALET processing of all Kepler light curves. In addition to the statistical analysis and comparison with the results of the Kepler team we shall present previously unknown planetary candidates.

**Acknowledgements:**

The German part of the Corot exoplanet detection efforts are funded by the Bundesministerium für Wirtschaft (BMWi), Berlin, via the German Space Agency DLR, Bonn, under grants 50QM1004 and 50QM1401. The authors  express their deep appreciation for the continuous and still continuing discussions with the German CoRoT team: J. Cabrera and Sz. Csizmadia (DLR Berlin-Adlershof), E. Günther and A. Hatzes (Landessternwarte Tautenburg), H. Rauer (DLR Berlin-Adlershof), G. Wuchterl (Landessternwarte Tautenburg).  We thank the CoRoT Exoplanet Science Team (CEST) for many years of an exciting collaboration.

## 10.   Figure Captions

**Figure 1:** The haar wavelet function is shown (also known as Daubechie wavelet function of first order). This is the simplest type of a wavelet functions. This function shows the typical characteristics of the wavelet function.

**Figure 2:** Illustration showing the Stationary Wavelet Transform Denoising (SWTD). The upper left panel shows a quadshirp function overlaid with white noise. SWTD is able to reduce noise by reconstructing a noise reduced signal. The signal is low pass filtered several times using wavelet analysis (lower panels a5, d5-d1). Each stage a new noise level is selected (dotted line) and higher resolution wavelets are used to analyze the remaining signal (lower left panels). Finally the single noise filtered parts (lower right panels) are combined to recreate the denoised signal (upper right panel).

**Figure 3:** Monochromatic light curve (CoRoT-ID 0658589071 LRc08 E2_3744) monitored by the CoRoT space telescope. The telescope was pointed 79 days at the star field LRc08 in the direction of the galaxy center. The collected raw flux is plotted against time in Barycentric Julian Date (BJD). Panel (a): original light curve; the many discontinuities and jumps visible in the light curve are caused by high energy particle impacts on the CCD. Panel (b): reconstructed light curve after VARLET filtering. Panel (c): residual light curve after subtracting the reconstructed curve (panel b) from the original light curve (panel a). No discontinuities or artifacts are visible after subtraction. Panel (d): The residual light curve (panel c) was searched by EXOTRANS. A transit was detected at a period of 6.2182 days.

**Figure 4:** Low cadence (sample period 30 minutes) Kepler light curve (KIC 4827723) covering 509 days of observation. Stellar variability is caused by star spots, pulsation, flaring etc. of periodic and non-periodic nature. Panel (a): Original light curve. Panel (b): Reconstructed light curve. Panel (c): residual light curve after subtraction of the reconstructed light curve (panel b) from the original light curve (panel a). The reduction of the stellar variability by a factor of 100 requires a change in scale. The transit feature is still contained in this residual light curve and shown in Figure 5.

**Figure 5:** Panel (a): The VARLET filtered reconstructed light curve from Figure 4 is phase-folded at the period of 7.239 days and binned (1000 bins). The transit is clearly identifiable. Panel (b): Same phase-folded light curve but not VARLET filtered. The transit of the planetary candidate is barely identifiable if phase-folded at the correct period of 7.239 days (panel a). Various stellar variabilities lower the SNR.

**Figure 6:** panel (a): phase diagram of the Kepler stellar binary KIC 4067549 from the unfiltered light curve. Strong primary and secondary transits at a period of 7.6989 days are identified. Panel (b): phase diagram of the same light curve at the correct transit period after VARLET filtering. The depths of the transit events are altered by the filter process. In general, the direct determination of the relative transit

depth to calculate the relative radius of the binary or planet is not feasible. Panel (c): same as panel (b) but at a smaller scale by one order of magnitude. The positions of the transit ingress or egress (dashed vertical lines) are not affected by the VARLET filtering compared to panel (a).

**Figure 7:** panel (a): Kepler light curve KIC 4474645 phase-folded after VARLET filtering at the detected period of the eclipsing binary (3.89 days) and binned (300 bins). The stellar binary prevents the detection of additional (planetary) transits. Panel (b): The residual light curve after PHALET filtering and phase-folded at the period of the binary and binned. The binary is fully extracted from the light curve.

**Figure 8:** Panel (a): original light curve of KIC 11401755 which is also known as the Kepler-36 multi-planet system. Panel (b): reconstructed light curve. Panel (c) residual light curve which is searched for transits by EXOTRANS. Panel (d): phase diagram of the detected transit of Kepler-36c at the period of 16.23 days. Panel (e): phase diagram of Kepler-36c after PHALET filtering. The transit of Kepler-36c is fully removed. The remaining light curve is now searched again by EXOTRANS. Panel (f): phase diagram at a period of 13.852 days. The transit of Kepler-36b is clearly detected by EXOTRANS.

**Figure 9:** panel (a) periodogram in the range from 0.5 days to 20 days of the Kepler-36 light curve searched by EXOTRANS after VARLET filtering. Kepler-36c was detected easily at a period of 16.23 days and shows the strongest peak. Other strong peaks are the partial harmonics of this transit. Kepler-36b, the second planet in this system, shows only a very weak line at 13.825 days. This line is also masked by a 6:7 harmonic from Kepler-36c because both planets are in near resonant orbits. Panel (b): periodogram of the Kepler-36 light curve after PHALET filtering and searched again by EXOTRANS. The residual light curve was filtered by PHALET removing the transits of Kepler-36c. The peak at a period of 16.23 days from the transit of Kepler-36c and all partial harmonics are removed. The peak of Kepler-36b at a period of 13.825 days is not affected by PHALET and now easily detected.

**Figure 10:** The percentage of detected planets with a period of 1.51 days against the cumulative SDE-distribution for the respective planets is shown. The cumulative distribution of the SDE values is added vice versa. As a result the percentage of the detected planets above a SDE limit can be read off directly. The percentage is set in proportion to the total number of synthetic generated light curves with planets. The vertical line shows the SDE threshold calculated out of the control group.

# Figures

**Figure 1:**

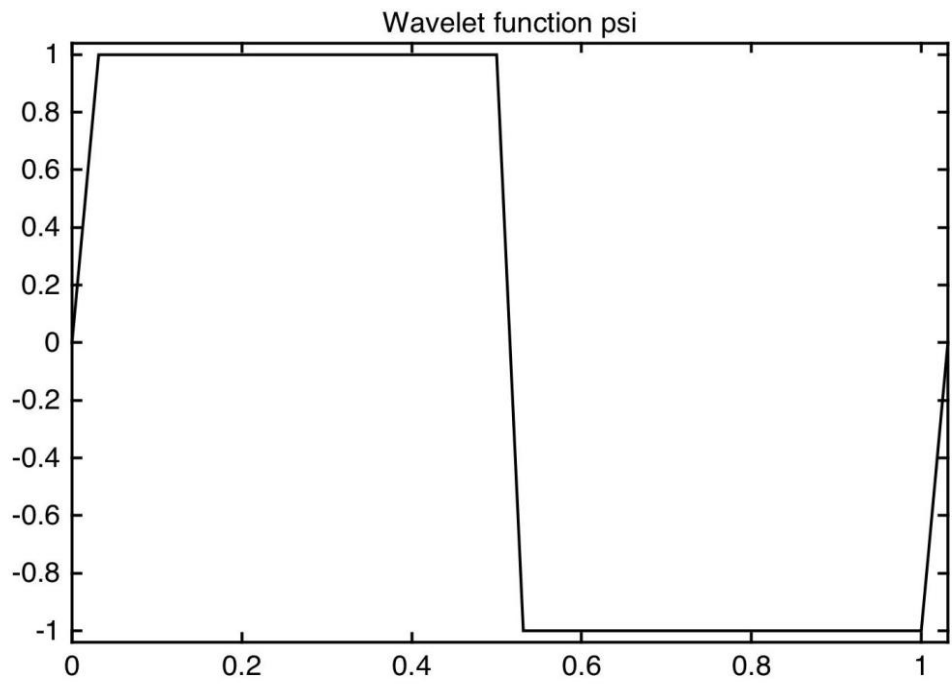

**Figure 2:**

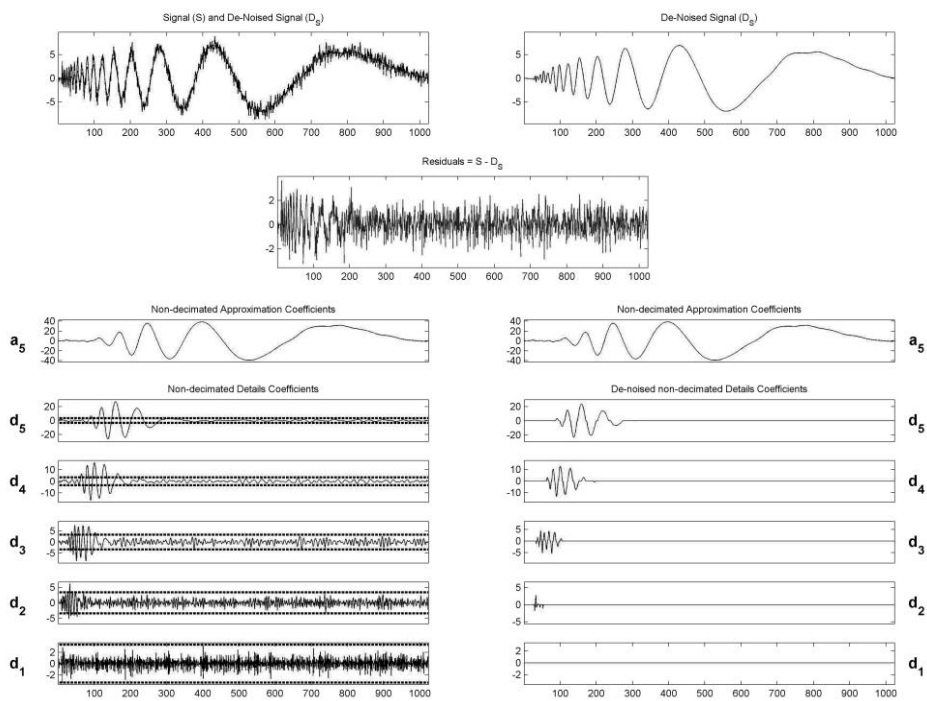

**Figure** **3:**

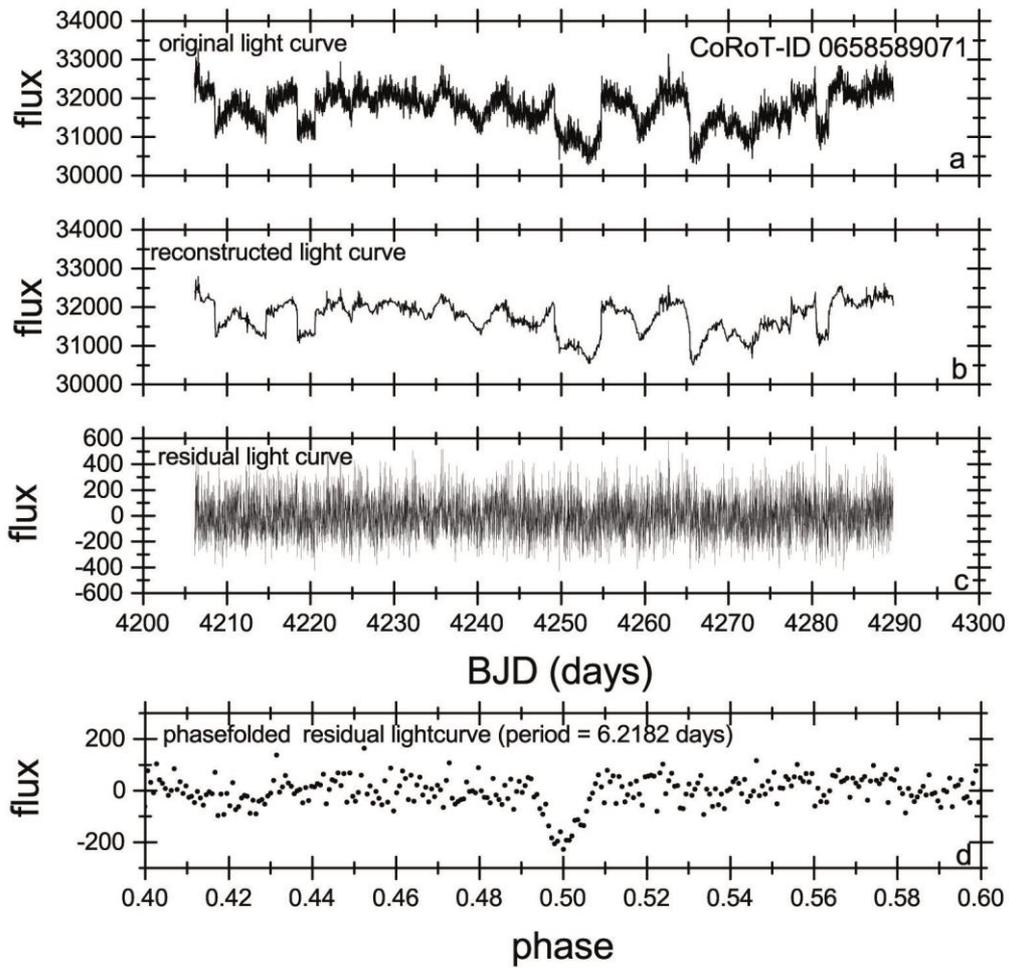

**Figure 4:**

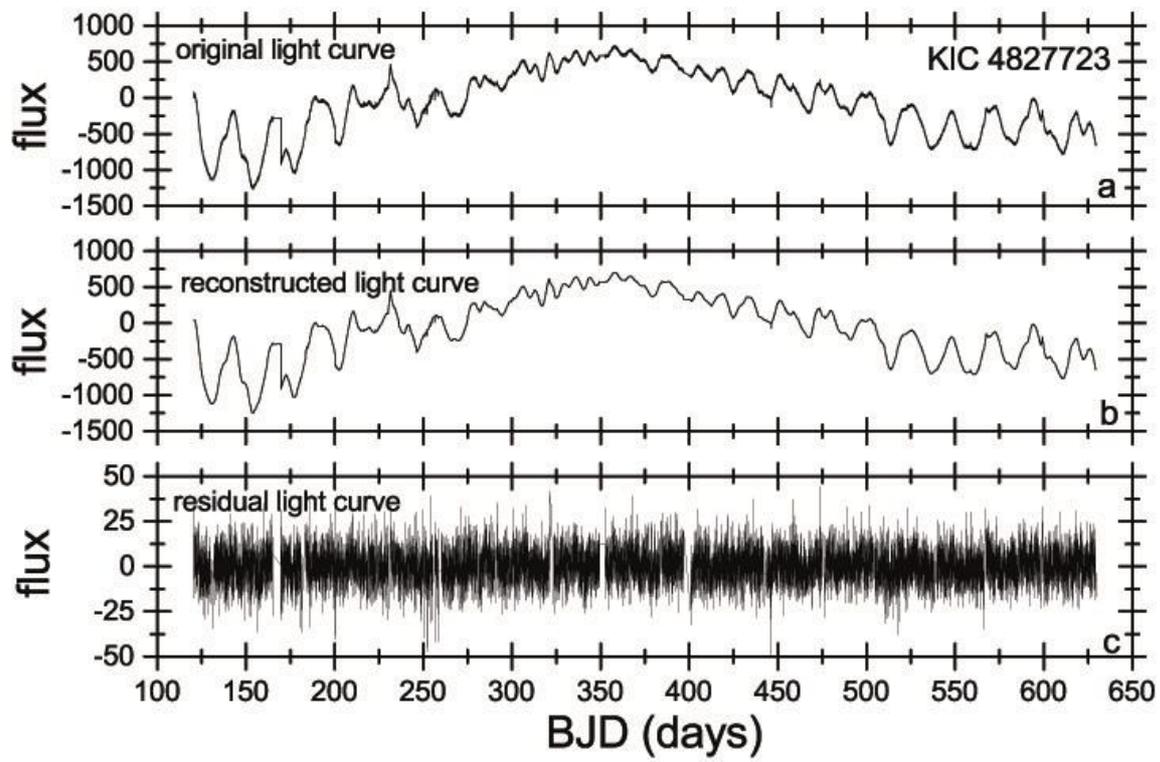

**Figure 5:**

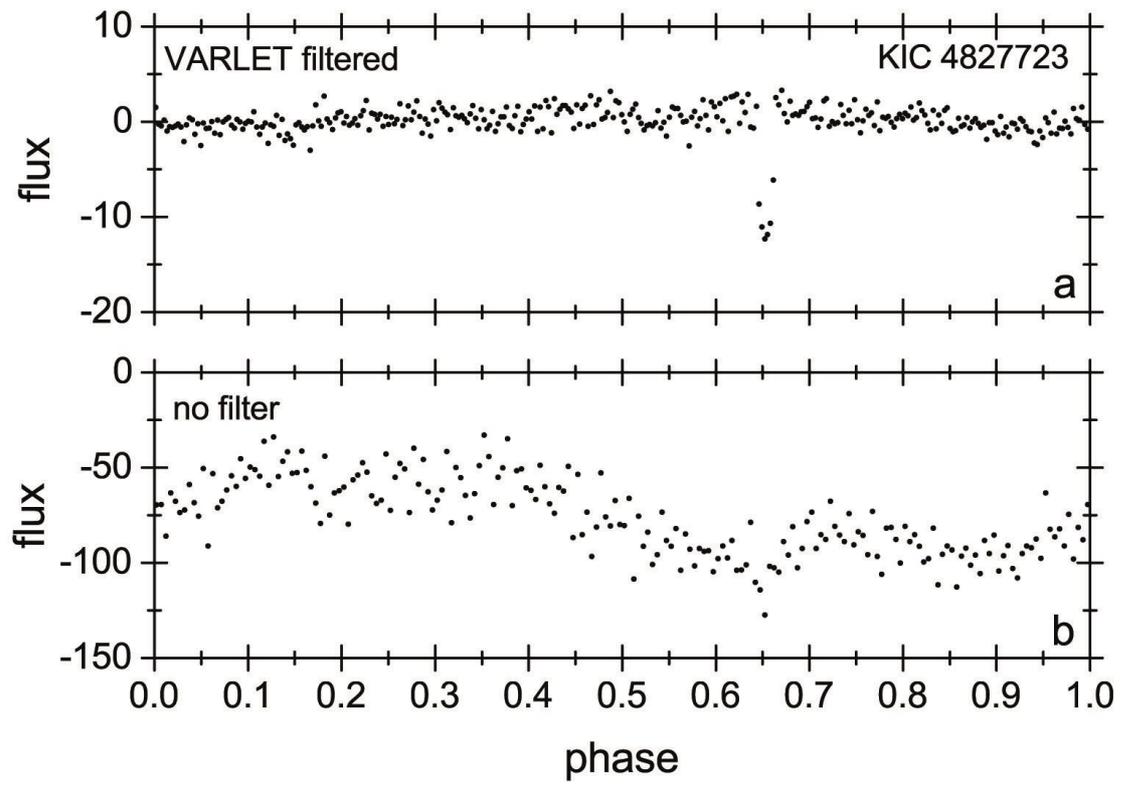

**Figure 6:**

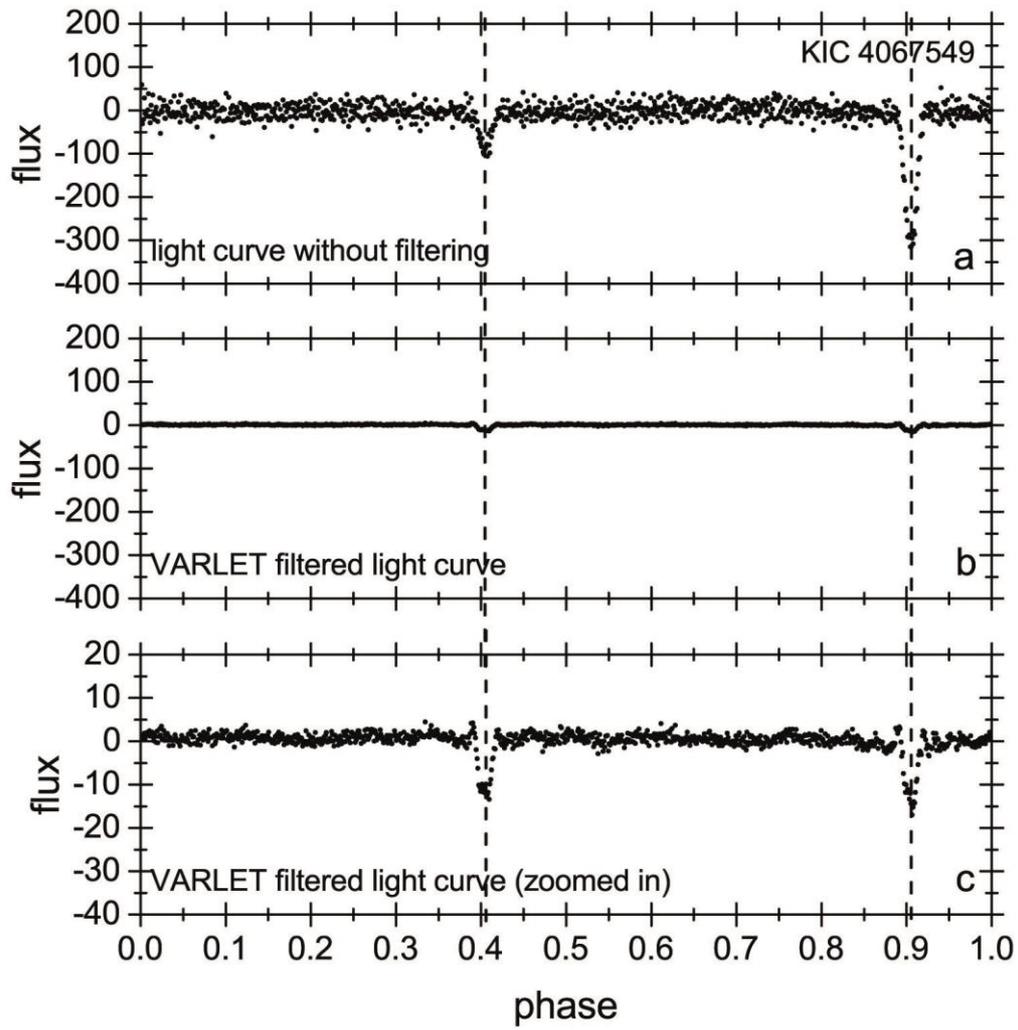

**Figure 7:**

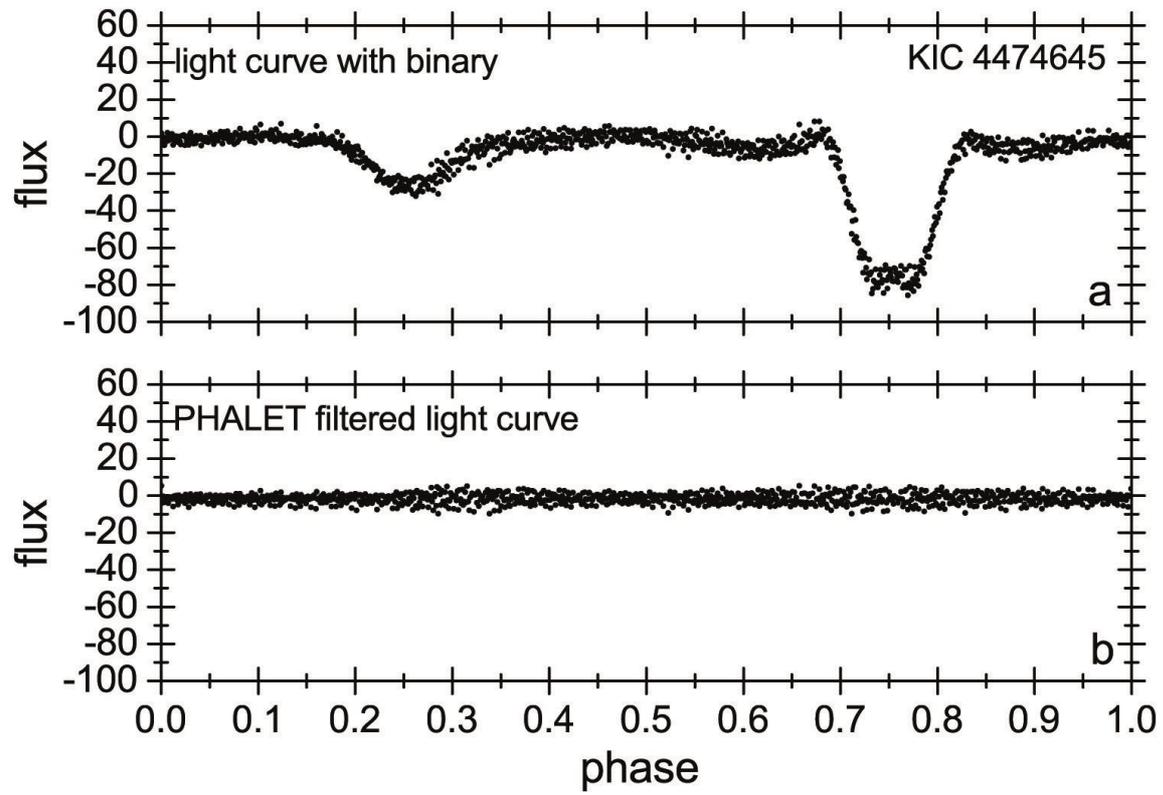

**Figure 8:**

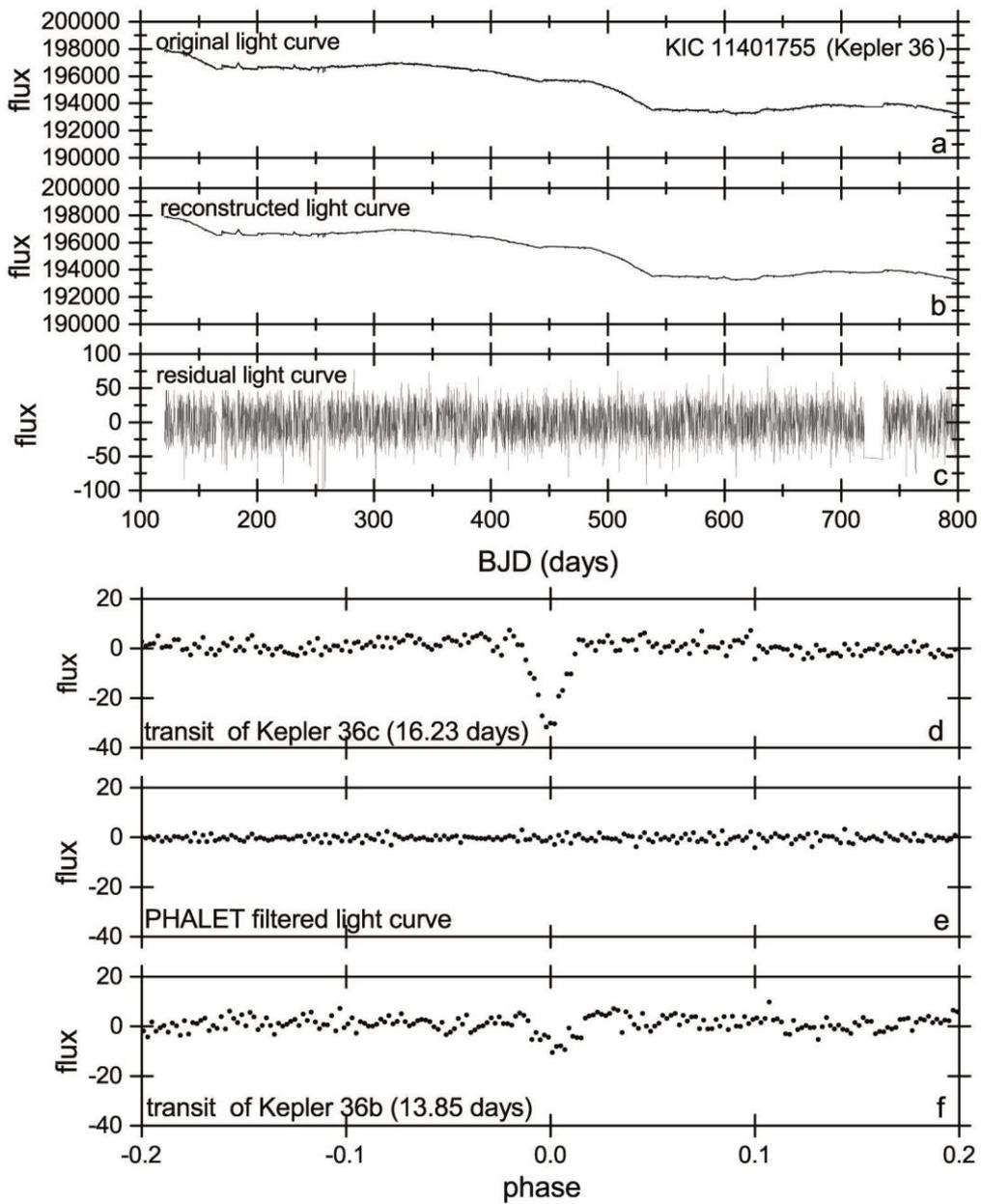

**Figure 9:**

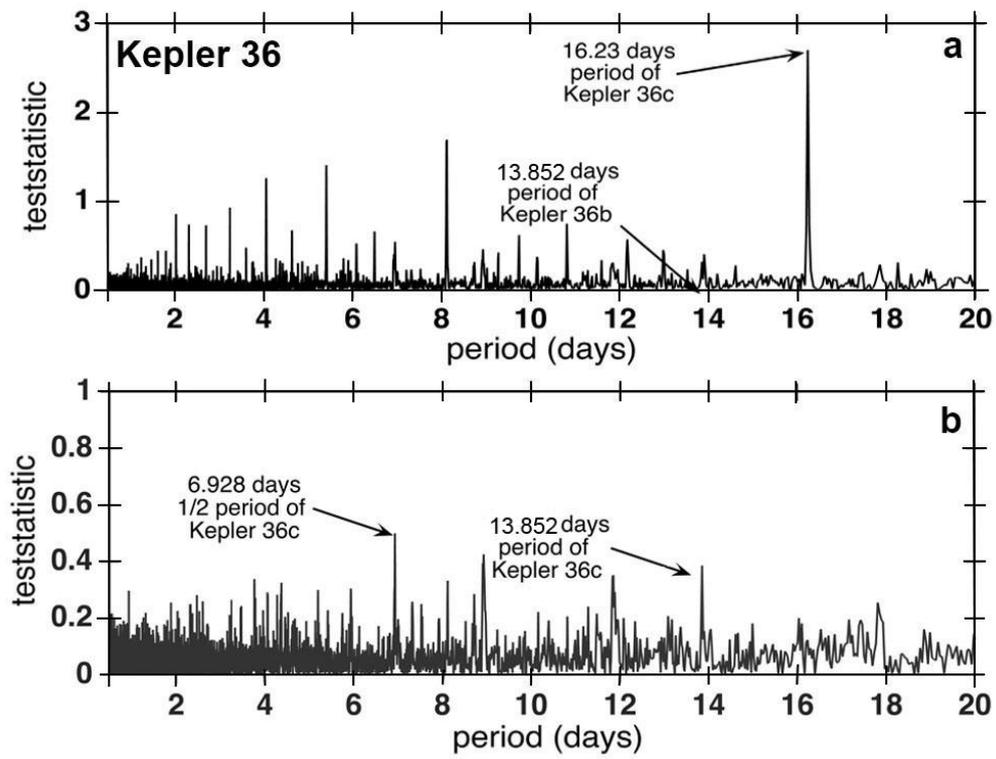

Figure 10:

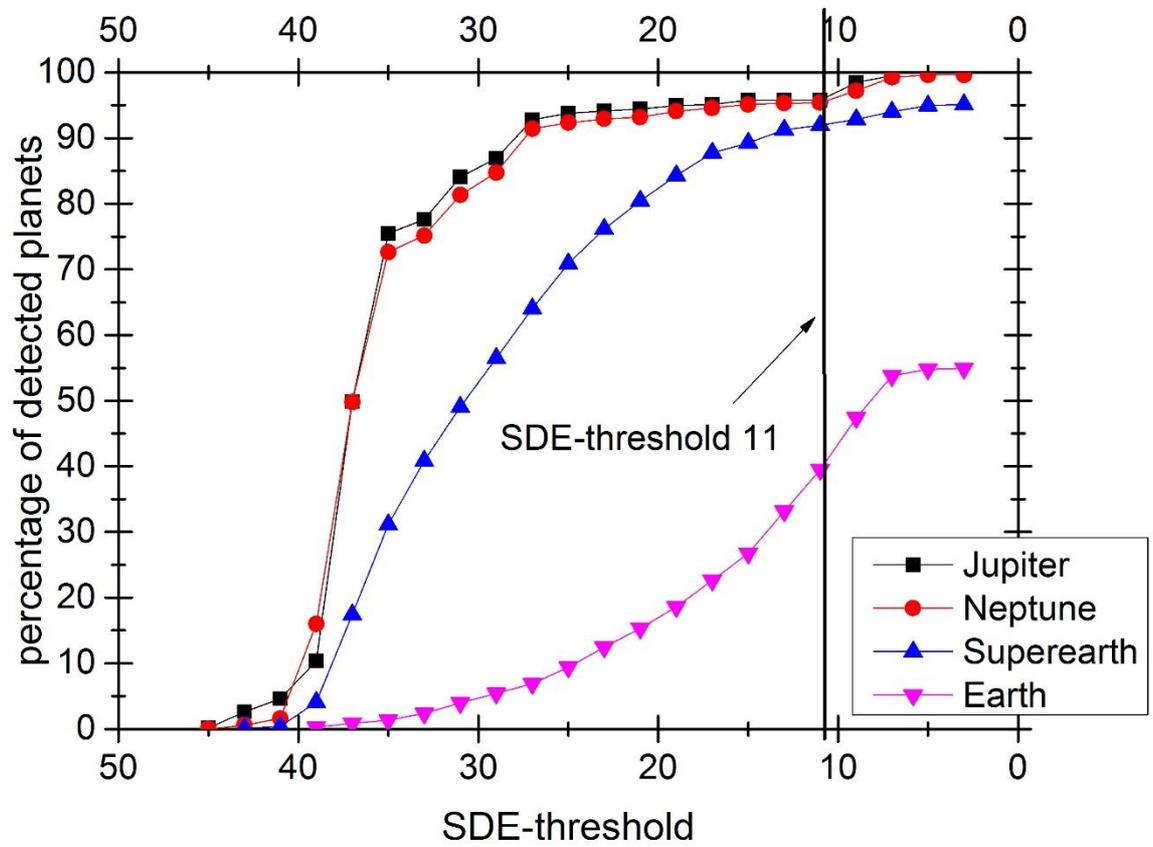

**Table 1:** number and percentage of detected planetary candidates by EXOTRANS after VARLET filtering.

| planetary radius (solar radii) | orbital periods of simulated planetary candidates | | | |
|---|---|---|---|---|
| | 1.51 days | 5.13 days | 10.31 days | 19.26 days |
| 0.01 (Earth) | 1021 (55%) | 441 (24%) | 69 (4%) | 88 (5%) |
| 0.02 (super-Earth) | 1766 (95%) | 1560 (84%) | 1321 (71%) | 843 (45%) |
| 0.04 (Neptune) | 1849 (99%) | 1781 (96%) | 1714 (92%) | 1545 (83%) |
| 0.1 (Jupiter) | 1854 (>99%) | 1825 (98%) | 1770 (95%) | 1707 (92%) |

**Table 2:** number and percentage of detected planetary candidates by EXOTRANS only.

| planetary radius (solar radii) | orbital periods of simulated planetary candidates | | | |
|---|---|---|---|---|
| | **1.51 days** | **5.13 days** | **10.31 days** | **19.26 days** |
| 0.01 (Earth) | 0 (0%) | 1 (<1%) | 0 (0%) | 1 (<1%) |
| 0.02 (super-Earth) | 4 (<1%) | 1 (<1%) | 1 (<1%) | 1 (<1%) |
| 0.04 (Neptune) | 77 (4%) | 20 (1%) | 5 (<1%) | 2 (<1%) |
| 0.1 (Jupiter) | 793 (43%) | 462 (25%) | 230 (12%) | 88 (5%) |